\documentclass[12pt,preprint]{aastex}

\begin{document}

\title{THE SMALLEST MASS RATIO YOUNG STAR SPECTROSCOPIC BINARIES}

\author{{L. PRATO\altaffilmark{1}, M. SIMON\altaffilmark{2},
T. MAZEH\altaffilmark{3}, \\ I. S. MCLEAN\altaffilmark{1},
D. NORMAN\altaffilmark{2}, AND S. ZUCKER\altaffilmark{3}}}

\altaffiltext{1}{Department of Physics and Astronomy, UCLA,
Los Angeles, CA 90095-1562; lprato@astro.ucla.edu}
\altaffiltext{2}{Department of Physics and Astronomy, SUNY,
Stony Brook, NY 11794-3800}
\altaffiltext{3}{Department of Physics and Astronomy, Tel Aviv University,
Tel Aviv, Israel}

\begin{abstract}

Using high resolution near-infrared spectroscopy with the Keck telescope, we 
have detected the radial velocity signatures of the cool secondary components 
in four optically identified pre$-$main-sequence, single-lined spectroscopic 
binaries. All are weak-lined T Tauri stars with well-defined center of mass
velocities. The mass ratio for one young binary, NTTS 160905$-$1859, is
M$_2$/M$_1$ $=$ 0.18$\pm$0.01,
the smallest yet measured dynamically for a pre$-$main-sequence 
spectroscopic binary.
These new results demonstrate the power of infrared spectroscopy for the 
dynamical identification of cool secondaries. Visible light spectroscopy, to 
date, has not revealed any pre$-$main-sequence secondary stars with masses
$<$0.5 M$_{\odot}$,
while two of the young systems reported here are in that range.
We compare our targets with a compilation of the published young
double-lined spectroscopic binaries and
discuss our unique contribution to this sample.

\end{abstract}


\section{Introduction}

The most important characteristic of a star is its mass.
Although there have been recent advances in determining the
masses of 1$-$2 M$_{\odot}$ pre$-$main-sequence (PMS) stars,
there exist only a few measurements of the masses of
sub-solar mass PMS
stars \citep{sim00b}.  As a result,
theoretical PMS evolutionary tracks have not
been well calibrated
at the low-mass end where uncertainties in the models
give rise to scatter as large as a factor of 3 in mass
and 10 in age \citep{sim00}.
Dynamical measurements of the masses of close binary systems
comprised of coeval components with a small mass
ratio provide an initial step towards tests of the PMS tracks over a broad
parameter range.

In the last decade, much progress has been made towards 
determining the masses of PMS stars. 
\citet{ghe95}, \citet{sim96}, and \citet{ste01}
take advantage of near-infrared (IR) speckle interferometry
and the fine guidance sensors
available on the {\it HST} to map the orbits of young binary systems.  
Using optical wavelength ($\sim$5000 \AA)
spectroscopy, \citet{mat89} and \citet{mat94}
identified twelve young, low-mass,
double-lined spectroscopic binaries
(SB2s) and thirteen single-lined systems (SB1s).
Following this seminal work,
other groups have recently detected new PMS
spectroscopic binary systems.  We discuss the sample of published SB2s 
in \S 5.

Our immediate goal is to measure mass
ratios and to enlarge the sample to study the mass ratio
distribution of young stars formed in diverse environments.
Precise measurements of the mass ratio distribution among young
star spectroscopic systems provide critical input for
theories of star formation.  For example, \citet{bat00} shows how
the density profiles of progenitor molecular cloud cores
determine the distribution of mass ratios in the close ($<$ 1 AU) binary
population which forms there.  Absolute masses
can be obtained from SB2s which are eclipsing or from SB2s with
known orbital elements determined from high resolution mapping
of the binary orbits (e.g., Boden et al. 1999).

SB2s identified by optical wavelength observations tend to have mass
ratios close to one.  As first
described in \citet{pra98} and further developed
in \citet{maz01}, our approach to detecting low mass ratio systems is
based on the strategy that, at IR wavelengths, the
flux of the secondary star is a greater fraction of the total
flux than at optical wavelengths.
For example, for two blackbodies of temperatures 5000
and 3000 K, the ratio of the Planck functions at 5000 \AA~ is
about 50, whereas at 1.555 $\mu$m, the ratio is about 4.
\citet{maz01} demonstrate the efficacy of this
technique, applied to three main-sequence SB1s.  \citet{maz97}
used an intermediate approach, detecting
main-sequence binaries
with mass ratios of $0.57\pm0.02$ and $0.48\pm0.03$ in the $R$ band.
With the advent of powerful cross-dispersed echelle spectrometers, such as 
NIRSPEC on the Keck telescope, the IR approach has become even more 
efficient.

In this paper we describe new observations of four PMS SB1s 
for which we have detected the low mass companion by cross-correlation 
against an extensive new library of spectral type standard star templates. 
The observations and data reduction methods are described in \S 2, the 
template spectra are presented in \S 3 and the derived dynamical mass ratios 
are discussed in \S 4.  Section 5 provides a discussion of the results, 
including a comparison with the complete sample of previously published SB2s 
and a comparison of our mass ratios with the predictions of evolutionary 
tracks in the H$-$R diagram.  A summary appears in \S 6.
Previous reports on applications of this
technique have appeared in \citet{pra98}, \citet{ste01},
and \citet{maz01}.

\section{Observations and Data Reduction}

\subsection{Observations}

\subsubsection{NIRSPEC Observations}

Spectra of the target objects were
obtained in 2000 June and July and in 2001 January and June; specific
dates and target properties appear in Table 1.  We used NIRSPEC,
the W. M. Keck Observatory's cross-dispersed
near-IR cryogenic spectrometer \citep{mcl98, mcl00}.
NIRSPEC's spectroscopic detector is a 1024 $\times$ 1024 ALADDIN III InSb
array that is sensitive from 0.95$-$5.5 $\mu$m.  In
high resolution mode, NIRSPEC
yielded a measured spectral resolution of R $=$ 24,000
in 9 orders across the $H$ band, centered
at $\lambda = 1.555 \mu$m in order 49.
Experience has shown that this spectral region contains
lines useful for the characterization of stars ranging from
spectral types of early G through late M, and is reasonably
free from terrestrial absorption lines \citep{pra98, maz01}.

The plate scale is 0.$''$144/pixel in the dispersion direction, and
0.$''$193/pixel in the cross-dispersion direction, and the two
pixel slit width was 0.$''$288.  A 15$''$
nod along the slit enabled us to obtain sequential spectra
at two positions on the detector for background
subtraction.  NIRSPEC's slit-viewing camera allowed
simultaneous monitoring and correction of the object's
position on the slit during the spectroscopic exposure. 
Integration times for a single exposure varied
from 60 to 600 s, depending on the source brightness.
Multiple flat and dark frames were
median filtered and the results differenced and normalized
to create a final flat field.

Internal Ne, Ar, Xe and Kr arc lamps usually provided wavelength scale
calibration and dispersion solutions.   OH night sky lines, with
identifications from \citet{rou00}, were also used.
A comparison of the
night sky line and arc lamp line calibrations for the 2000
July data showed no difference, but the differences in the
2001 June data between the two calibrations was as much as 5
pixels.  This drift of the grating position over time 
on that occasion was attributable
to vibration caused by the image rotator mechanism inside of NIRSPEC.
Observations in the adaptive optics
(AO) mode (\S 2.1.2) were never affected by grating position drifts
since the AO system provides external compensation for field rotation
and NIRSPEC's internal rotator is not used.

\subsubsection{NIRSPEC Adaptive Optics Observations}

Because NTTS 155913$-$2233 is a hierarchical triple, with a
tertiary component located at a 0.$''$29 separation \citep{ghe93}, the
2000 June observations were taken with NIRSPEC behind the Keck II
facility AO system \citep{wiz00}.
Magnification by the AO reimaging optics alters the plate
scale, resulting in pixels a factor of 10.6 times
smaller.  Correspondingly, the two pixel slit
width becomes 0.$''$027.  The measured spectral
resolution behind the AO system is higher by about 25 \%,
yielding $R = 30,000$.
The NTTS 155913$-$2233 tertiary was easily resolved
with no overlap of the components' FWHM $=$ 0.$''$06
point spread functions.
High resolution
images, taken immediately before or after the
spectroscopic integrations,
indicated that the Strehl ratio was typically 0.25
in the $H$ band.
In AO mode, the slow input beam and plane parallel window produce a 
Fabry-Perot type fringing pattern which contaminates the observed spectra. A 
notch or Hanning filter successfully removed most of this artifact during 
data reduction.

\subsection{Data Reduction}

Two approaches were used for the data reduction.
At Stony Brook, we extracted the spectra using software
originally written by \citet{pra98} and adapted to the format 
of the NIRSPEC spectra.  At UCLA, spectra were extracted using software
developed by S. S. Kim, L. Prato, and I. McLean.  Comparison
of extractions of the same object by the two different approaches yielded
spectra that were indistiguishable; the locations of the spectral
lines were identical to within $\sim 1$ km s$^{-1}$.
Order 49 had no detectable terrestrial
absorption lines, as expected from prior experience and inspection of
the infrared solar atlas \citep{liv91} and
order 50 displayed only a few weak terrestrial lines.
Unfortunately, orders 51, 52, and 53 contained so many strong absorption
lines as to be unusable. The atmospheric contamination in orders 45$-$48 
was intermediate between these extremes.  We used only
the spectra in order 49, and confirmation from the spectra in
order 50, for the results reported in this paper.

\bigskip
\section{Template Spectra}

To search for the presence of the secondary in the target spectra,
an extensive new suite of template standard star
spectra were required.  Table 2 lists the 23 main-sequence stars
observed for this purpose.  The spectral type appears in column (2),
as provided by
the Centre de Donne\'es astronomiques de Strasbourg (SIMBAD).
Column (3) lists the Julian Date of the
observation and column (4) the derived
radial velocity \citep{maz01}.  All but 8 of the 
template stars have radial velocities measured by the 
Geneva \citep{duq91} or CfA groups \citep{lat85}. The average 
difference between the radial velocities we measured and those of
\citet{duq91} and \citet{lat85}
is $-0.2\pm0.8$ km s$^{-1}$.  We therefore adopt $\pm 1.0$ km s$^{-1}$
as an estimate of the systematic uncertainty between the  zero-points 
of the velocities we shall derive for the components of spectroscopic 
binaries we observed and their center-of-mass ($\gamma$) velocities
derived by others.  

The excellent sensitivity, resolution, and
spectral range obtained with NIRSPEC are demonstrated in Figure 1, where
the 23 $H$ band template spectra for order 49 are plotted
with corrections for heliocentric motion and radial velocity applied.  
The continuum has been flattened.
The correlation coefficient resulting from
cross-correlation of two template spectra against each other provides
a measure of the stars'
similarity.  The template spectra have been arranged, in Figure 1 and
in Table 2, in order of decreasing mutual correlation, from hottest
to coldest.  HD 283750 and GL 275.2A appear out of order with respect
to their spectral types, K2 and M4, respectively, as given in SIMBAD.
For HD 283750,
the discrepancy is only 3 spectral sub-classes,
however, GL 275.2A has been incorrectly classified
as an M4. Its IR spectrum is obviously that of a K star.

Based on Poisson statistics,
the signal to noise ratio of the spectra shown in Figure 1 is
typically greater than 100; most of the observed features are real.
The G through early M type standard star spectra are dominated by atomic
lines, predominantly Fe~I, Ni~I, Si~I, and Ti~I.  Beginning around
spectral type K5, OH lines appear and increase in depth. A
pronounced transition is evident between the M3 and M4 subclasses;
the earlier type standards generally show a clearly definable
continuum, whereas objects later than M3, dominated by water features,
do not display any region
clear of lines from which a continuum might be definitively
characterized.  In addition to H$_2$O, late M star spectra
in this region contain strong OH lines and the CO $\Delta v =$ 3
bandhead at $\sim$1.558 $\mu$m.

\bigskip
\section{Results}

We selected the targets for our study from the sample of
SB1s described by \citet{mat94} and here report on the
results for four systems,
NTTS 155913$-$2233, NTTS 160905$-$1859, Parenago 1771, and Parenago 1925;
observations of other systems are still in progress.  A summary
of the properties of the four target SBs appears in Table 1.
Each of these binaries was detected
as an X-ray source by the {\it Einstein} observatory
\citep{ku82, wal86, mat89, gio90}; all are weak-lined
T Tauri stars with deep
photospheric absorption lines.  The $\gamma$
velocities have already been well determined,
from optical observations, for these four systems \citep{mat94}.
Figure 2 shows representative order 49 spectra of the targets as 
observed without correction for heliocentric velocity.  The
continuum has been flattened for each spectrum.

We analyzed the spectra, following the procedure described in \citet{maz01},
with the TODCOR algorithm, which uses a
two-dimensional cross-correlation for the analysis of composite
spectra \citep{zuc94}.  Table 3 presents the results for each
target: columns (1) and (2) list, respectively, the Julian
date and the orbital phase of the observation.  
TODCOR analysis identifies the 
templates that provide the best fit to the primary and secondary, their 
velocities, $v_1$ and $v_2$, given in columns (3) and (4) of
Table 3, and an estimate of their flux ratio (see Table 4).
The $\gamma$ velocity is known for our targets
from prior analysis as SB1s \citep{mat94}, so the mass 
ratio of the components, $q=M_2/M_1$, column (5) of
Table 3, follows from their velocities 
relative to $\gamma$, $q = (v_1-\gamma)/(\gamma-v_2)$ \citep{maz01}.
Values for $\gamma$ appear at the top of each section of Table 3
adjacent to the target names.  For all 4 of our targets,
the velocities listed in Table 3 are
the averages of the analyses of order 49 using
different template stars.  Preliminary results from order 50 were
consistent with these.
The errors in the component star velocity determinations 
were propagated to determine a final uncertainty in the mass ratio.

\subsection{NTTS 155913$-$2233} 

This NTTS is a hierarchical triple, consisting of an SB with a period
of 2.42 days and a tertiary component located at 0.$''$29 separation
\citep{mat89, ghe93, wal94}.
Our observations of this multiple were obtained using the AO system to
separate the SB and the tertiary.
TODCOR analysis identified GL1094, rotationally broadened to 15 km s$^{-1}$,
as the best-fitting template for the primary.  Its spectral type, K5,
agrees with the determination of \citet{mat89} for the SB seen in visible
light (Table 1).  GL436, also rotationally
broadened to 15 km s$^{-1}$, provided the best
fit to the secondary, with a flux ratio of 0.39 $\pm$0.03
for two of the observations.  On
UT 2000 June 11, however, TODCOR identified a flux ratio of 0.79 $\pm$0.03,
although the mass ratio was identical.  Since NTTS 155913$-$2233 has
a relatively short period and the components are probably only
a few solar radii apart, the stars potentially fill their
Roche lobes \citep{mat89},
and hence are perhaps distorted, producing variation in their
flux ratio with phase.

The upper left two panels of Figure 3 are cuts through the
two-dimensional correlation function,
results of the 2000 June 6 data TODCOR analysis.  The 
primary and secondary correlation peaks, located
at $\sim +43$ and $\sim -74$ km s$^{-1}$, upper and lower panels,
are clearly seen.  The correlation plot for the secondary also shows
a residual
bump at $\sim +43$ km s$^{-1}$, the result of the imperfect match of the
primary template to the primary of NTTS 155913$-$2233. 
Table 3 lists the  component velocities measured in our three observations
of the system.  The derived mass ratio, using the $\gamma$ velocity 
measured by \citet{mat89}, is $0.63\pm0.02$.  This value is the average
of two independent observations of the system.  The value of q in
parenthesis has a large uncertainty because it was measured
when the component velocities were near the $\gamma$ velocity.
Therefore, we did not include it in the average.

\subsection{NTTS 160905$-$1859}

The maximum correlation was obtained for NTTS 160905$-$1859 using the K2 star
HD 283750, rotationally broadened
to 25 km s$^{-1}$, as the primary template. This spectral type is
consistent with the determination of \citet{mat89} for the primary.
For our highest signal to noise spectra, obtained without the 
AO system, the TODCOR correlation was maximized by using GL 402, an M4,
as the secondary template, also broadened
to 25 km s$^{-1}$.  An average flux ratio of $\sim 0.24$ was derived.
The range of flux ratios found with different templates and derived
from distinct observations was 0.15 $-$ 0.35.
Because of the lower throughput and hence signal to noise
ratio in the AO measurements, it was not possible to extract secondary
star velocities from these data.  The upper right panels of
Figure 3 show the results of
the correlation analysis of the UT 2001 June 2
data. Table 3 summarizes the measured velocities.  The ratio
derived using the
$\gamma$ velocity from \citet{mat89} is $0.18\pm0.01$, the lowest
yet measured for a PMS spectroscopic binary.  

\subsection{Parenago 1771} 

GL1094, rotationally broadened
to 15 km s$^{-1}$, gave the largest correlation as
the primary of Parenago 1771.  Its
designated spectral type is K5, in good agreement
with the K4 type given for the SB1
by \citet{mat94}.  GL436, broadened
to 15 km s$^{-1}$, with an average flux ratio of 0.26 with
respect to the primary, gave
the best correlation for the secondary.  For different templates,
the range of flux ratios was 0.22 $-$ 0.32.  The lower left hand plots in
Figure 3 show the correlation for the primary (upper panel)
and secondary using the order 49
spectra.  In the lower panel,
the correlation peaks for the secondary
at $\sim 64$ km s$^{-1}$ and for the residual of the
primary at $\sim 14$ km s$^{-1}$ are nearly equal in amplitude,
the result of an imperfect
match between the main-sequence primary template and the primary in
this PMS binary.  The derived
mass ratio is $0.34\pm0.03$.  It is
necessary to confirm this mass ratio because it is based on only one
observation.

\subsection{Parenago 1925} 

TODCOR analysis of the Parenago 1925 system
identified HD 283750 and GL669B, each 
rotationally broadened to 10 km s$^{-1}$ and with an
average flux ratio $\sim 0.23$, as 
the best fitting primary and secondary.  For different templates,
the range of flux ratios was 0.12 $-$ 0.35.
The assigned spectral type of 
HD 283750 is K2, in good agreement with the classification by \citet{mat94} 
of the primary as a K3.  The lowest right panel of
Figure 3 clearly shows the 
correlation maximum associated with the secondary, but also shows a 
comparable residual peak associated with the primary.
The mass ratio in this system is 
$0.31\pm0.04$.  As for Parenago 1771, since this result is based on only
one observation it must be confirmed.
            
\parindent=0.5in
\section{Discussion}


To illustrate the sensitivity of the IR technique, 
Figure 4 combines our sample with the IR detected SB2 NTTS 045251+3016 
\citep{ste01} and 26 optical wavelength detected SB2s.  From Figure 4
it is evident that the small sample presented 
in this paper includes the lowest mass ratios ever measured for PMS SB2s.  
Table 4 lists the entire sample of PMS SB2s plotted in Figure 4, including 
the location of the binaries, column (2), their mass ratios, 
periods, flux ratios,
and primary masses (see below),
in columns (3), (4) (5), and (6), and the references used 
for these data, column (7).  

Figure 4 displays the well-known bias of SB2s detected in visible
light toward mass ratios of one.  Our results start to fill in the low
mass ratio part of the observed distribution.  A concentrated effort to
observe a complete sample of SBs will be required to determine the true
distribution of PMS SB2 mass ratios.
It is interesting to notice (Table 4) that the five SB2s
contributed so far by the IR technique span the same
range of periods identified
among the optically detected
SB2s.  A complete sample of SB2s will therefore enable tests of
theoretical predictions of the mass ratio distribution as a function
of binary separation (e.g., Bate and Bonnell 1997).

High angular resolution
imaging studies of young binaries reveal a broad range of 
component mass ratios, down to the regime of brown dwarfs, as inferred 
photometrically by comparing the location of the binary components in the 
H$-$R diagram with PMS evolutionary
tracks \citep{whi99, whi01, woi01}.  However,
differences among the theoretical calculations of
PMS evolution produce considerable scatter among the inferred 
mass ratios (e.g., Figure 7 of Woitas et al. 2001).  Dynamical measurements 
offer the means to determine the mass ratios and their
distribution precisely and thus to test specific predictions of calculations
of binary formation (e.g., Bate 2000).   

Except for the eclipsing systems, indicated in Table 4, and 
the spatially resolved NTTS 0455251+3016 \citep{ste01}, we
cannot derive dynamical values for the component masses of the
SB2s in Figure 4 because their orbital inclinations are not yet known.
In order to compare our IR SB2
secondary masses with their minimum
values set by the mass function,
and to learn the mass range currently
sampled by dynamical techniques, we estimate
the primary masses, and thus M$_2$
using the dynamical mass ratios
(Table 5).  The procedure we use to obtain M$_1$ is
described in detail in Appendix A.  The M$_1$ values for the entire SB2
sample are included in Table 4, column (6).

For the 4 IR SB2s described in this paper,
columns (3), (4), and (7) of Table 5 provide estimates for M$_1$, M$_2$, and
the minimum mass, M$_{2_{min}}$, derived
by setting the orbital inclination to $\pi/2$ in the mass function f(M)
of \citet{mat94}, column (6).
The uncertainties in M$_2$ and M$_{2_{min}}$ include the
uncertainties in the mass ratio and in our estimate of M$_1$.
In all cases, M$_2$ is consistently
greater than its predicted minimum value.

In Figure 5 we plot mass ratio as a function of estimated M$_1$.
Figure 5 shows that smaller secondary masses,
as well as smaller mass ratios, are more readily 
identified with IR spectroscopy;  all the companions detected
by the IR technique have M$_2$ $<$ 1 M$_{\odot}$.
In general, few primary star masses less than 1 M$_{\odot}$
and no primary star masses less than 0.7 M$_{\odot}$ have been
detected in PMS SB2s.  This is illustrated by the gap on the left
hand side of Figure 5 and is probably the result of a selection
effect for brighter systems among SB surveys.
The total sample shown in Figure 5 is weighted
towards PMS stars with little or no circumstellar material,
although spanning a wide range of ages, from $\sim$ 10$^5$ yr through
$\sim$ 10$^7$ yr.  Young stars which lack
circumstellar envelopes and disks do not experience veiling
of their photospheric absorption lines, and hence provide
easier targets for the identification of spectroscopic binaries.

The essential result of TODCOR analysis is the mass ratio of the
components.  This value is robust because it is determined by
cross-correlation of the many spectral lines in common between the 
target and templates.  The analysis also provides estimates of
the primary and secondary spectral types and their $H$ band flux ratio,
which give the maximum correlation with the target.  The templates
almost certainly do not match exactly
to the spectra of the stars in the target binaries
because the surface gravities of PMS stars are
lower than those of main-sequence stars, and because
the templates and the PMS targets may differ in metallicity.
Similarly, $H$ band flux ratios do not necessarily represent
accurate component luminosity ratios.
Nonetheless, it is of interest to investigate the extent to
which the values of $M_2/M_1$ implied by our dynamical results are
consistent with PMS evolutionary tracks in the H$-$R diagram,
presented in Figure 6.

Our approach to estimating
the effective temperature, T$_{eff}$, and luminosity for each object,
and the associated uncertainties, is described in Appendix B.
We use the tracks of Palla \& Stahler (1997) because they are representative
of PMS evolution, similar to the results of Baraffe et al. (1998) and
Siess et al. (2000), and span the range of masses of our sample objects.
Figure 6 shows that, to within 1 $\sigma$, the secondary
masses implied by the mass ratios
(Table 5) are consistent with the secondary masses estimated from
the tracks (column (5) of Table 5), except for Parenago 1925.
Only for NTTS 160905$-$1859 do the components lie on distinct
isochrones; the other systems appear to be coeval to 1 $\sigma$.
It will be possible to derive absolute, dynamical masses 
for these objects within the next decade as ground and space based
interferometers become available for mapping their orbits.
These mass determinations will be precise to within a few
percent or less, a requirement for meaningful testing of models.
Reliable luminosity estimates, however, cannot be obtained
by dynamical means and will require not only spatially
resolved observations, but also multi-wavelength
data in order to obtain precise values.

\section{Summary}

We sucessfully observed 4 PMS SB1 systems using IR spectroscopy to find
cool companion stars not detected at optical wavelengths. A suite
of spectral type standard stars provided templates for 
cross-correlation analysis and yielded a spectral library
of early G through late M type stars.
We have identified the smallest mass ratio PMS SB2s yet
observed, demonstrating the power of this approach.
The binary components we have studied cover a large
mass range, from 2 M$_{\odot}$ primaries to 0.2 M$_{\odot}$
secondaries.

To provide a comparison to our sample and a useful reference,
we compiled as complete a list as possible
of PMS SB2s and their main properties.
We estimate primary masses of all known PMS SBs (Table 4) and 
qualitatively use these values, 
together with the precise mass ratios, in Figure 5
to illustrate the extremely low secondary star masses
discovered with our IR observations.
In addition, we place both components of the PMS SBs
systems studied in our sample on the H$-$R diagram to compare
the evolutionary tracks with our dynamical results.
To within 1 $\sigma$, most of these results are consistent, however,
mass estimates based on placing the
stars on the H$-$R diagram are still approximate because the
spectral types and luminosities of the component stars are
incompletely characterized.

The application of optical wavelength spectrometers
on 1$-$2 m class telescopes for the identification of SB1s, to be
subsequently examined in the IR, is a practical technique for the
measurement of large samples of PMS SB2s.  Although such a study
is magnitude limited, and thus does not sample the smallest
mass primaries, the detection of SBs ({\it not} SB2s)
is unbiased with respect to the mass ratio.  We anticipate that, 
for a given star forming region, a complete survey of SBs at optical
wavelengths with IR follow up for large mass ratio systems would provide 
the true mass ratio distribution for the smallest separation
observable systems, to the limiting magnitude of the initial 
optical light survey.  Because of the multiple epochs of
observations required, as well as the limited time available
at large telescopes, progress on this research is inevitably
slow-paced.  As we proceed with new confirmed observations,
we plan to publish these results as they become available.

\bigskip
\section{Acknowledgements}

We thank the staff, observing assistants, and support 
scientists of the W. M. Keck Observatory for their logistical and technical
help.  We are grateful to R. Mathieu for providing orbital data in advance
of publication, to C. Bender for producing Figure 1, to C. McCabe
for helpful discussions, and to an anonymous referee for suggestions
which improved our presentation.
This research has made use of the SIMBAD database, operated
at CDS, Strasbourg, France.  This research was supported in part
by NSF Grant 98-19694 (to MS), and by US-Israel Binational Science
Foundation Grant No. 97-00460 and the Israel Science Foundation (to
TM).  Data presented herein were obtained at the W.M. Keck 
Observatory, which is operated as a scientific partnership
between the California 
Institute of Technology, the University of California and NASA.
The Observatory was made possible by the generous financial support of the 
W.M. Keck Foundation.
The authors wish to extend special thanks to those
of Hawaiian ancestry on whose sacred mountain we are
privileged to be guests.

\appendix

\section{Primary Mass Determinations for Figure 5}

Primary masses for fourteen of the binaries in Table 4 were
found in the literature (column 7).  Among these, only for
NTTS 045251+3016 \citep{ste01} is there a published value
for the flux ratio.
To obtain the primary masses for the remaining systems, 
we did the following.  The total luminosity of the SB2 was obtained
from the literature or calculated
from the total $K$ band magnitude, as described in
\citet{sim95}.  The primary luminosity was then determined from
the flux ratio, when available.  Otherwise, if
the mass ratio of an SB2 was within 3\% of unity,
a component bolometric luminosity
ratio of 1.0 was assumed.  For the remaining three SB2s, we used an
approximate flux ratio based on the empirical mass ratio $-$
$\Delta$m$_K$ relation shown in Figure 5 of \citet{whi01}.

For cases in which no component spectral type data
was available, we assumed that
the spectral type of the unresolved system was equal to
the T$_{eff}$ of the primary.  For the 4 systems described in this paper, 
we estimated T$_{eff}$ from the best fit primary star template (\S 4)
and  the spectral type to temperature conversion
found in \citet{luh00} (see Appendix B).

Primary star masses were derived from the tracks of \citet{pal99}
and \citet{bar98} for the lower mass objects.  Objects plotted
on both sets of tracks usually had comparable masses to within 10 \%;
for consistency we used \citet{pal99} for all estimates.  We estimate 
that the uncertainties in the primary masses determined from
the H$-$R diagram and listed in Table 4 are about 10\%.

\section{T$_{eff}$ and Luminosity Estimates for Figure 6}

Spectral types for the 8 individual stars in our SBs sample
were determined from the spectral types of the templates which
gave the maximum correlation.  Because of the caveats in the
accuracy of this approach (see \S 5) we assigned an uncertainty
to these results of two spectral subclasses for secondary
stars.  In all cases the primary star spectral types derived
from the TODCOR analysis were the same as previous estimates of
the spectral types for these stars, or else within one spectral subclass.
Therefore, we assigned a one subclass uncertainty
to the primary star spectral types.  Typically the
same component velocities were determined using a range of
template spectral types.  To convert to T$_{eff}$, we used
Figure 5 of Luhman (2000).

To derive component luminosities
we used the 1.555 $\mu$m flux ratio found by TODCOR and
applied this proportionately to the total luminosity.  
For the PMS stars of the age and mass range considered here,
their $H$ band flux is approximately proportional to their
luminosity.  We estimate one source of uncertainties from
the range of flux ratios produced by
TODCOR for different templates (see \S 4).  For example,
for Parenago 1771, an M3 template gave the best correlation for
the secondary star, yielding a velocity of 64.5 $\pm$1.5 km s$^{-1}$
(Table 3).  An M7 template yielded only a 1 \% difference in velocity,
yet a $\sim$25 \% difference in the flux ratio.
Thus, although the determination of the velocity is robust,
the characterization of the flux and hence luminosity ratio is 
uncertain.  Because these error estimates are approximate,
we use a $\sim$15 \% distance uncertainty, or else the uncertainty
produced by the range in flux ratios, whichever is larger.

Other sources of error might arise from the proximity
of the stars in very short period SBs, which may distort 
their structure.  The light ratio variation (\S 4.1) detected in
NTTS 155913$-$2233 (period $=$ 2.4 days) may indicate this effect. Such 
systems are likely to be less useful for tests of the PMS 
evolutionary tracks.

\clearpage

\begin{figure}
\plotone{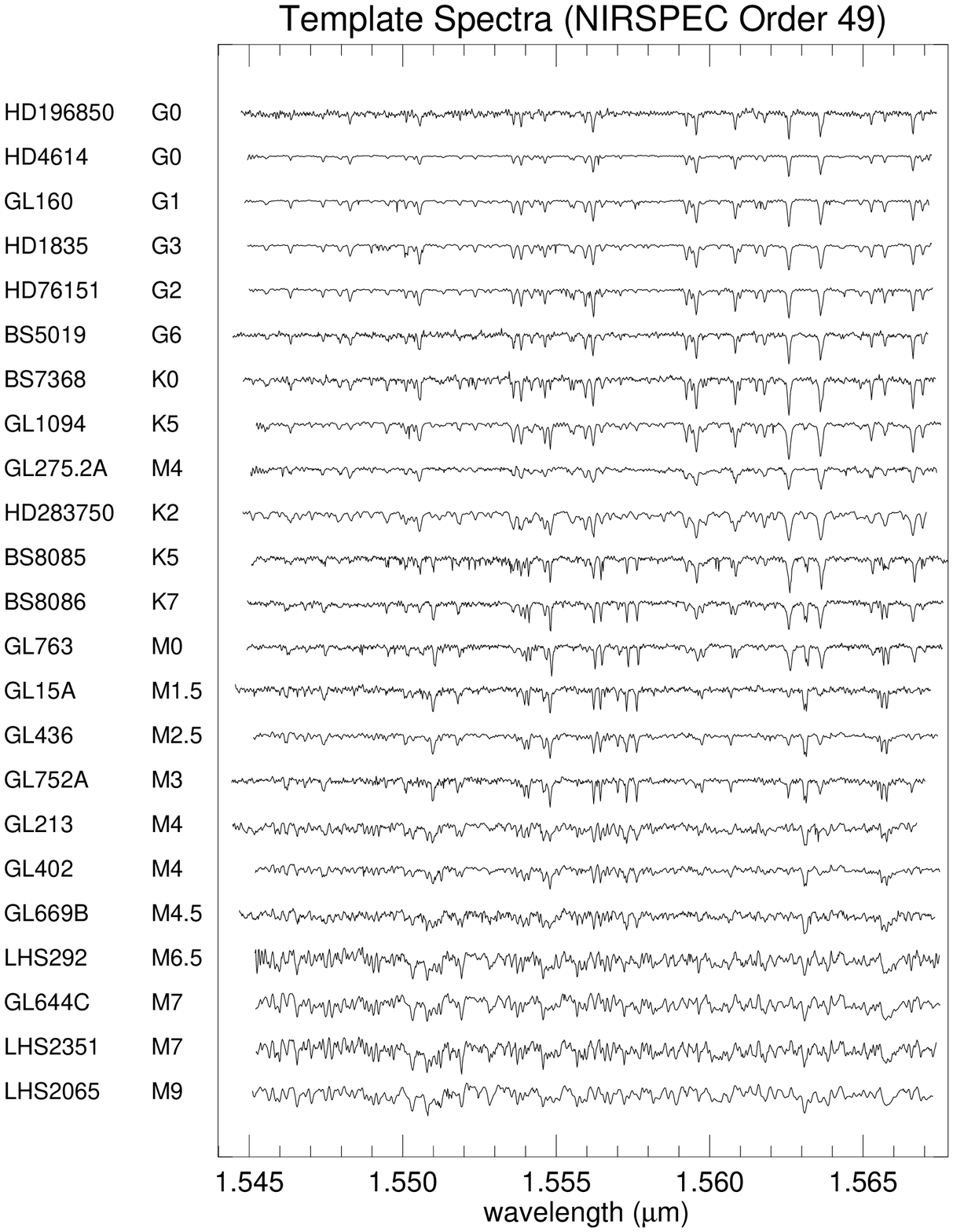}
\caption{The NIRSPEC template spectra for order 49.
The spectral continuum has been flattened for each template.  \label{fig1}}
\end{figure}

\begin{figure}
\plotone{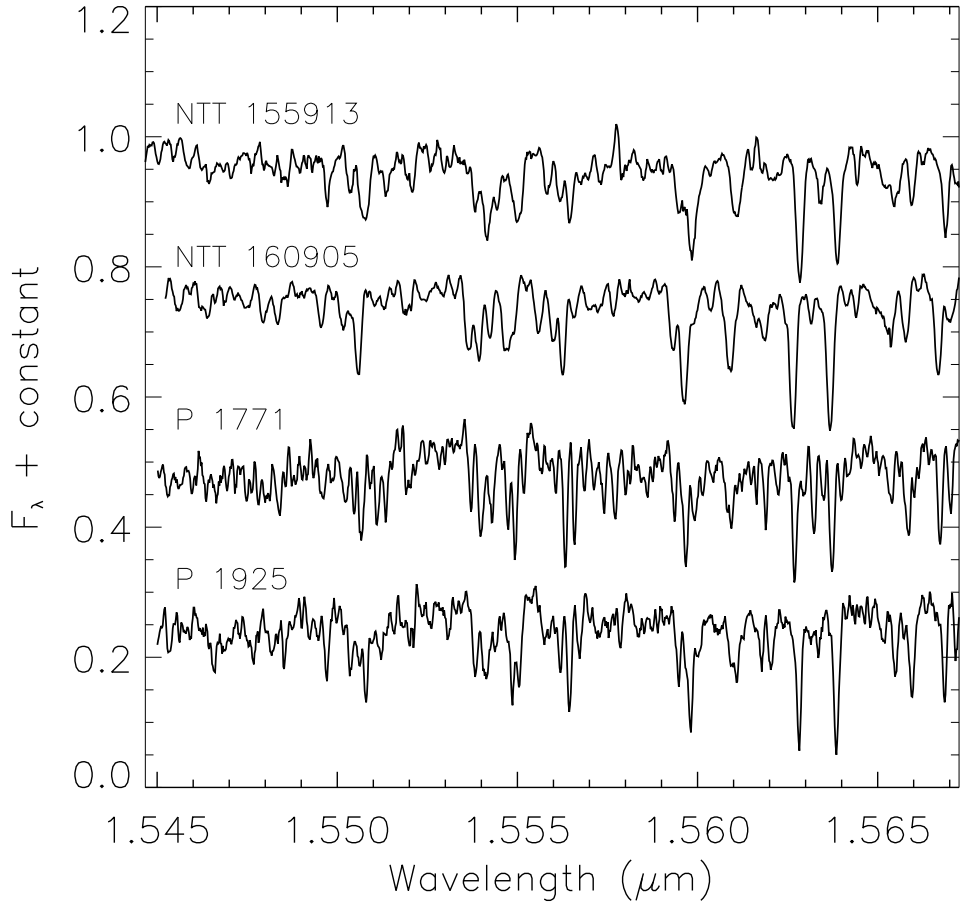}
\medskip
\caption{Order 49 NIRSPEC spectra of
NTTS 155913$-$2233, NTTS 160905$-$1859, Parenago 1771, and
Parenago 1925, observed respectively
on UT 2000 June 6 (AO), UT 2001 June 1, and, for the latter two, UT 2001
January 5 (AO).  The Parenago 1771 and Parenago 1925 spectra have been smoothed
with a two-pixel boxcar. No heliocentric or radial velocity
corrections have been applied. 
The spectral continuum has been flattened for each object.\label{fig2}}
\end{figure}

\begin{figure}
\plotone{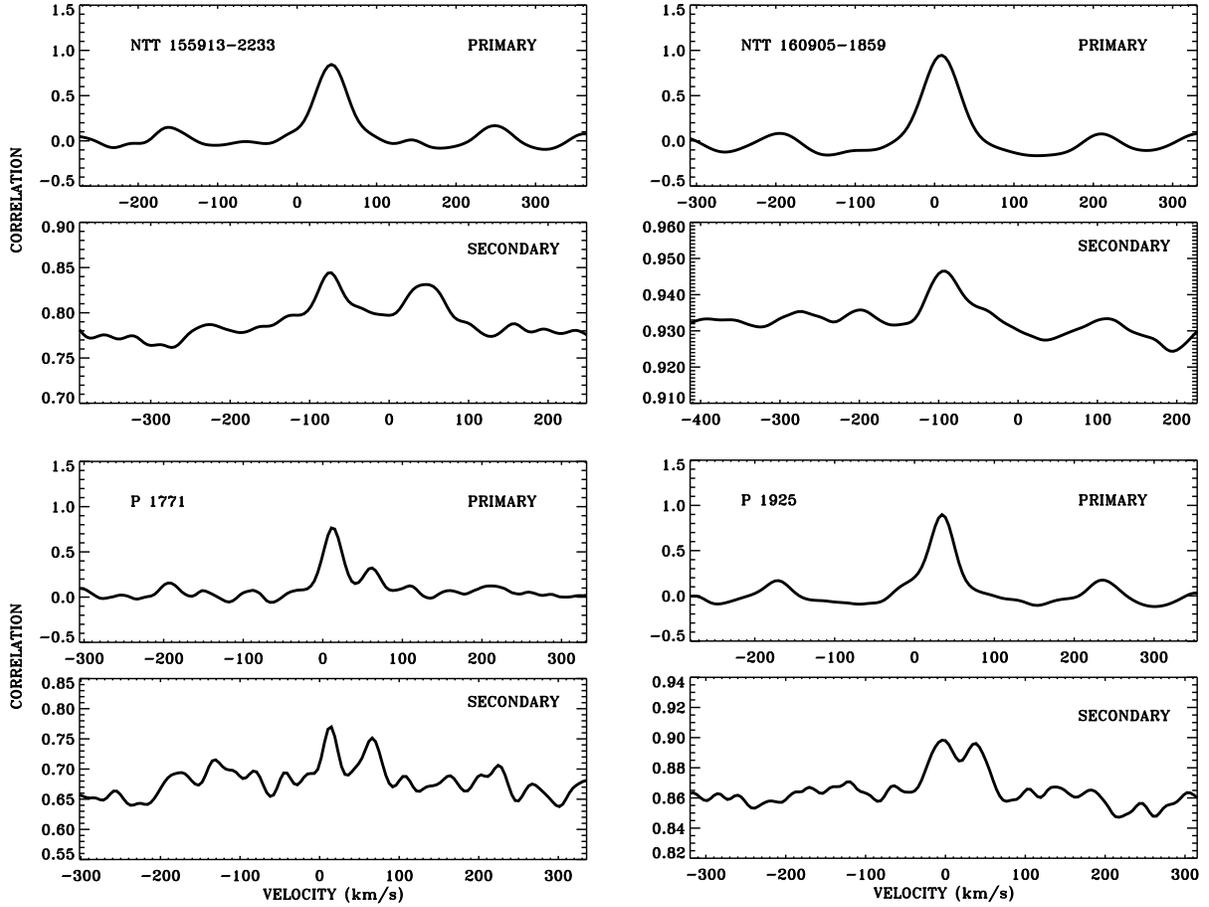}
\caption{TODCOR correlation results for the spectra in Figure 2.  For each
binary, the upper panel shows the two-dimensional cross-correlation as a 
function of the radial velocity of the best fitting primary template,
holding the secondary velocity fixed. The lower panel shows the
cross-correlation as a function of the secondary template velocity.
  \label{fig3}}
\end{figure}

\begin{figure}
\plotone{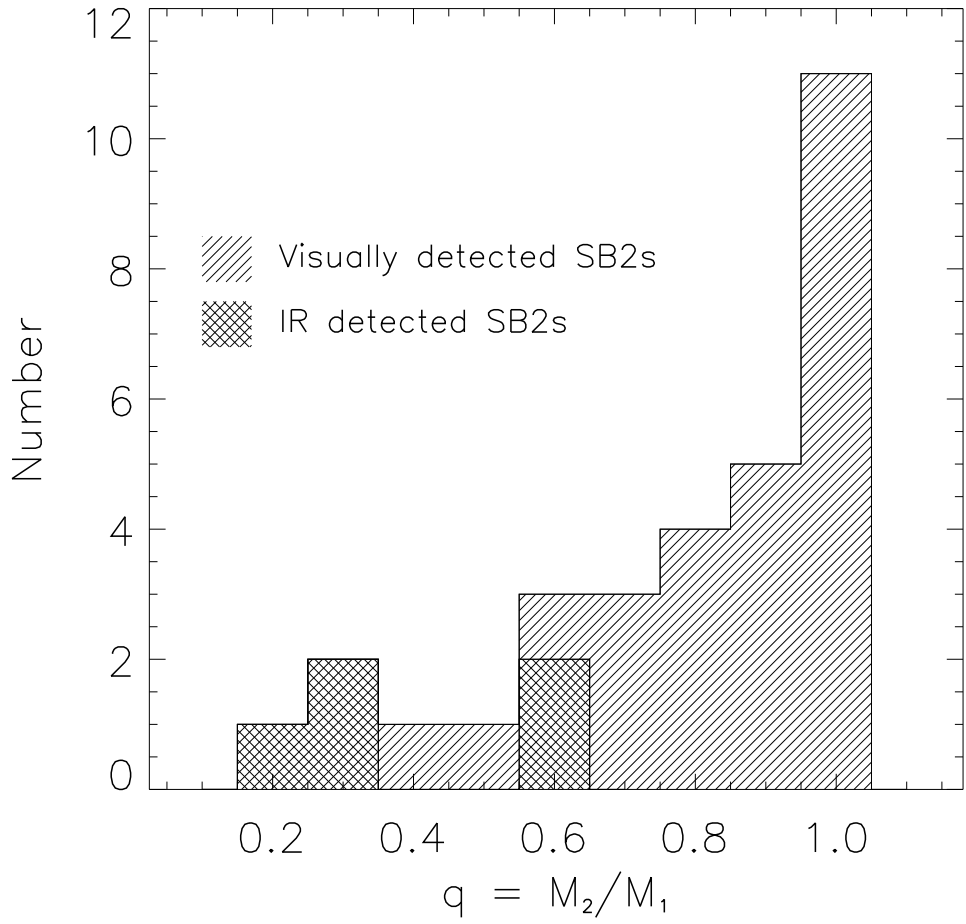}
\medskip
\caption{Histogram of currently known SB2 mass ratios. The 
systems shown, their mass ratios, and other data are
listed in Table 4.  \label{fig4}}
\end{figure}

\begin{figure}
\plotone{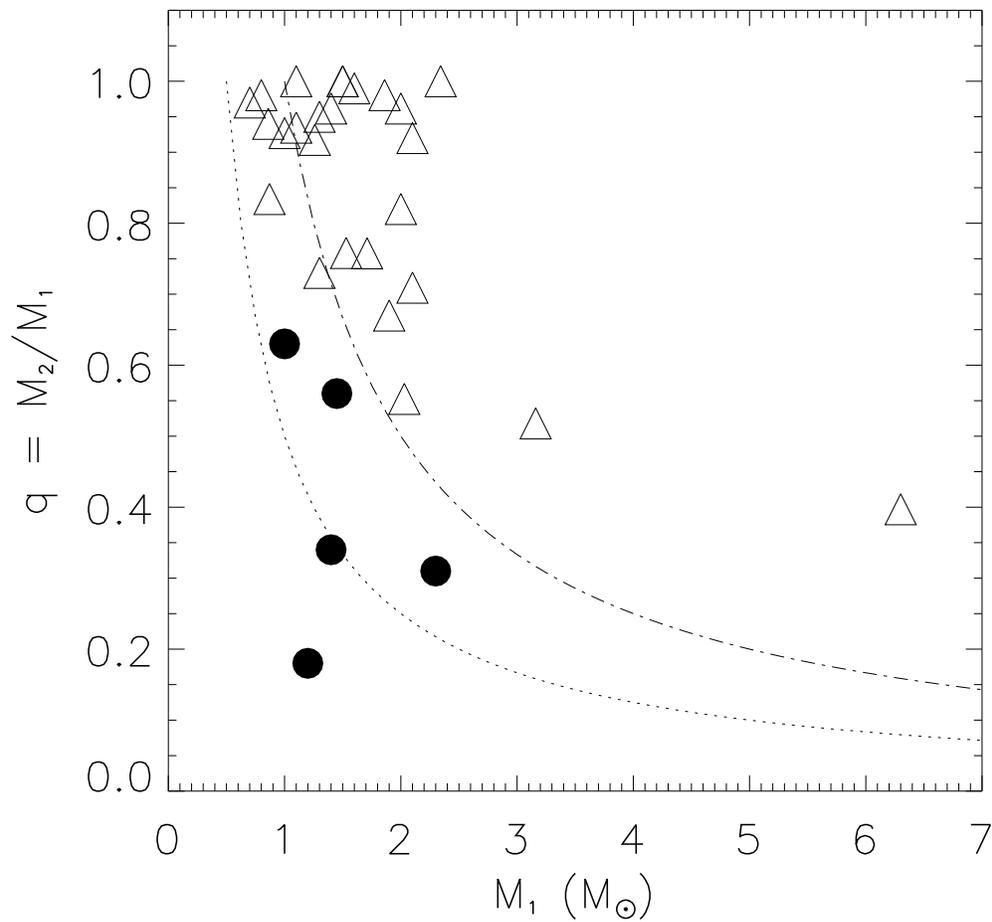}
\medskip
\caption{Mass ratios as a function of primary star mass.
Triangles are SB2s discovered at optical wavelengths and filled circles are
SB2s discovered in the IR.  Primary masses were taken from the
literature or calculated as described in the text, \S 5.
The dot-dash line shows the curve
for a constant M$_2 = $1.0 M$_{\odot}$ and the dotted line for
a constant M$_2 = $0.5 M$_{\odot}$.  \label{fig5}}
\end{figure}

\begin{figure}
\plotone{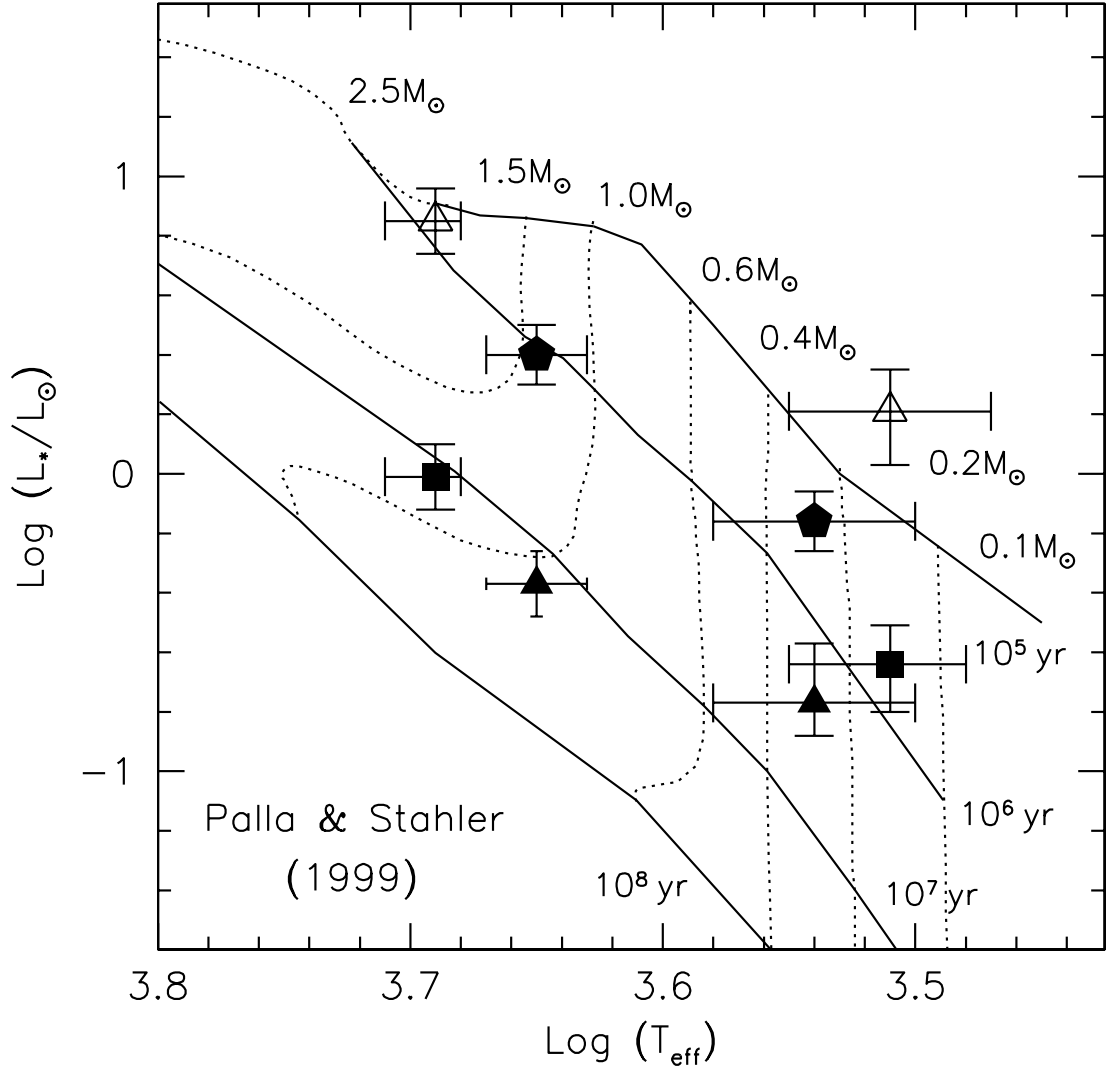}
\medskip
\caption{Both components of the target systems plotted on the
H$-$R diagram, with the evolutionary tracks of \citet{pal99}
superimposed.  The plotting symbols are assigned as
follows:  open triangles, Parenago 1925; pentagons, Parenago
1771; squares, NTTS 160905$-$1859; filled triangles,
NTTS 155913$-$2233.  \label{fig6}}
\end{figure}

\clearpage

\pagestyle{empty}

\begin{deluxetable}{lcccccc}
\rotate
\tablewidth{0pt}
\tablecaption{Summary of Sample Properties and Observations \label{tbl-1}}
\tablehead{
\colhead{$ $}   & \colhead{R.A. (2000.0)} & \colhead{Dec. (2000.0)} &
\colhead{H$_{total}$} & \colhead{Spectral} &
\colhead{Period} & \colhead{UT Date and Mode}\\
\colhead{Object} & \colhead{(${h},{m},{s}$)} & \colhead{(${\circ},',''$)} &
\colhead{(mag)\tablenotemark{a}} & 
\colhead{Type\tablenotemark{a}} & \colhead{(days)\tablenotemark{a}} & 
\colhead{of observation}}
\startdata
Parenago 1771 &  05 35 09.75 & $-$05 23 26.8 & 9.6 & K4 & 149.5 & 2001 January 5 (AO) \\
Parenago 1925 &  05 35 18.34 & $-$05 22 37.7 & 8.6 & K3 & 32.94 & 2001 January 5 (AO) \\
NTTS 155913$-$2233 & 16 02 10.5 & $-$22 41 29 &   8.2 & K5 & 2.42378 & 2000 June 7, 10, 11 (AO)\\
NTTS 160905$-$1859 & 16 11 59.33 & $-$19 06 52.40 & 8.2 & K1 & 10.400 & 2000 June 9, 11 (AO)\\
$ $ & $ $ & $ $ & $ $ & $ $ & $ $ & 2000 July 14 (non-AO) \\
$ $ & $ $ & $ $ & $ $ & $ $ & $ $ & 2001 June 1, 2 (non-AO) \\
\enddata

\tablenotetext{a}{Data from \citet{mat89}.}

\end{deluxetable}

\clearpage

\pagestyle{empty}

\begin{deluxetable}{lccr}
\tablewidth{0pt}
\tablecaption{Template Library \label{tbl-2}}
\tablehead{
\colhead{$ $} & \colhead{Spectral} & \colhead{Date (JD)} & \colhead{v$_{rad}$} \\
\colhead{Object} & \colhead{Type\tablenotemark{a}} & \colhead{2450000+} & \colhead{km s$^{-1}$}}
\startdata
HD 196850 & G0  & 1715.05  & -21.7   \\
HD 4614   & G0  & 1917.75  & 8.3    \\
GL 160    & G1  & 1917.77  & 23.6   \\
HD 1835   & G3  & 1917.72  & -2.5   \\
HD 76151  & G2  & 1917.03  & 31.8  \\
BS 5019   & G5  & 1714.84  & -9.0 \\
BS 7368   & G8  & 1715.04  & -22.3 \\
GL 1094   & K5  & 1917.86  & -30.9 \\
GL 275.2A & M4  & 1917.08  & -3.9 \\
HD 283750 & K2  & 1916.90  & 39.6 \\
BS 8085   & K5  & 1703.03  & -67.1 \\
BS 8086   & K7  & 1707.13  & -65.0 \\
GL 763    & M0  & 1715.11  & -60.5 \\
GL 15A    & M1.5& 1707.14  & 12.2 \\
GL 436    & M2.5& 1915.14  & 10.0\\
GL 752A   & M3  & 1705.03  & 37.4 \\
GL 213    & M4  & 1917.96  & 106.6 \\
GL 402    & M4  & 1915.04  & -1.1 \\
GL 669B   & M4.5& 1705.99  & -33.8\\
LHS 292   & M6.5& 1915.06  & 1.3 \\
GL 644C   & M7  & 1943.17  & 15.0\\
LHS 2351  & M7  & 1942.97  & -14.4 \\
LHS 2065  & M9  & 1942.94  & -8.4  \\
\enddata

\tablenotetext{a}{As given in the SIMBAD database.}

\end{deluxetable}

\clearpage

\pagestyle{empty}

\begin{deluxetable}{ccccc}
\tablewidth{0pt}
\tablecaption{Mass-Ratio Determination \label{tbl-3}}
\tablehead{
\colhead{JD} & \colhead{$ $} & \colhead{$v_1$} & \colhead{$v_2$} &\colhead{$ $} \\
\colhead{(2450000+)} & \colhead{Phase} & \colhead{(km s$^{-1}$)} & \colhead{(km s$^{-1}$)}  & \colhead{q}}
\startdata
\multicolumn{5}{c}{~~~~}\\
\multicolumn{5}{c}{Parenago 1771~~  $\gamma= 25.6\pm0.2$\, km s$^{-1}$}\\
  1914.86      &   ...  & $12.5\pm1.0$ &$64.5\pm1.5$& $0.34\pm0.03$\\
\tableline
\multicolumn{5}{c}{~~~~}\\
\multicolumn{5}{c}{Parenago 1925~~ $\gamma= 25.9\pm0.3$\, km s$^{-1}$}\\  
  1914.94      &  ...  &$34.7\pm1.0$&$-2.2\pm1.5$ & $0.31\pm 0.04$\\
\tableline
\multicolumn{5}{c}{~~~~}\\
\multicolumn{5}{c}{NTTS 155913-2233~~ $\gamma=-2.3\pm 0.7$ km s$^{-1}$ }   \\
  1702.96      & 0.22 & $43.4\pm1.1$ &$-74.4\pm1.5$ & $0.63\pm0.02$\\ 
  1705.93      & 0.45 & $-41.8\pm1.1$&$61.3\pm1.6$  & $0.62\pm0.03$\\ 
  1706.92      & 0.86 & $0.4\pm1.0$  &$-5.8\pm1.1$  & $(0.77\pm0.45)\tablenotemark{a}$\\ 
               &      &             &         &$\bar{q}=0.63\pm0.02  $ \\ 
\tableline
\multicolumn{5}{c}{~~~~}\\
\multicolumn{5}{c}{NTTS 160905-1859~~ $\gamma=-6.4\pm 0.2$ km s$^{-1}$ }   \\
  1704.91      & 0.73  &$-22.3\pm1.1$  &   ...   &     ...          \\ 
  1706.94      & 0.93  &$-15.6\pm1.1$  &   ...   &      ...         \\ 
  1739.89      & 0.10  &$8.4\pm1.0$    &$-93.3\pm2.1$& $0.17\pm0.01$\\ 
  2061.95      & 0.07  &$7.0\pm1.0$    &$-79.7\pm3.2$& $0.18\pm0.02$\\
  2062.91      & 0.16  &$10.8\pm1.0$   &$-99.3\pm2.3$& $0.19\pm0.02$\\ 
               &      &               &         &$\bar{q}=0.18\pm0.01  $ \\ 
\enddata

\tablenotetext{a}{Not used in $\bar{q}$.  See \S 4.1.}

\end{deluxetable}

\clearpage

\pagestyle{empty}

\begin{deluxetable}{lcccccl}
\tablewidth{0pt}
\tablecaption{PMS Double-Lined Spectroscopic Binaries \label{tbl-4}}
\tablehead{
\colhead{Object} & \colhead{Location} & \colhead{q} & \colhead{P (days)}& \colhead{Flux Ratio} & \colhead{M$_1$ (M$_{\odot}$)} & \colhead{Refs\tablenotemark{a}}}
\startdata
RX J0350.5$-$1355 & Orion & 0.92 & 9.28 & 0.82 & 2.1 & 1 \\
V773 Tau & Tau-Aur  & 0.76 & 51.08 & ...  & 1.53 & 2 \\
V826 Tau & Tau-Aur  & 0.98 & 3.91 & $\sim$ 1  & 0.8 & 3, 4 \\
RX J0441.0$-$0839 & Orion & 0.82 & 13.56 & 0.70 & 2.0 & 1 \\
DQ Tau & Tau-Aur  & 0.97 & 15.8 &  $\sim$ 1 & 0.7 & 5 \\
NTTS 045251+3016 & Tau-Aur & 0.56 & 2530 & 0.40\tablenotemark{b} & 1.45 & 6 \\
OriNTT 429 & Orion Belt & 1.0 & 7.46 & $\sim$ 1 & 1.5 & 4, 7, 8 \\
OriNTT 569 & Orion Belt & 1.0 & 4.25 & $\sim$ 1 & 1.1 & 4, 7, 8 \\
RX J0529.4+0041\tablenotemark{c} & Orion & 3.04 & 0.73 & ... & 1.3 & 9 \\
RX J0530.7$-$0434 & Orion & 1.00 & 40.57 & 1.00 & 1.5 & 1 \\
RX J0532.1$-$0732 & Orion & 0.95 & 46.85 & 0.82 & 1.3 & 1 \\
Parenago 1540 & Trapezium & 0.76 & 33.73 & ... & 1.71 & 2 \\
Parenago 1771 & Trapezium & 0.34 & 149.5 & 0.26\tablenotemark{b} & 1.4 & 4, 7, 10 \\
BM Ori\tablenotemark{c} & Trapezium & 0.40 & 6.5 & ... & 6.3 & 2 \\
Parenago 1925 & Trapezium & 0.31 & 32.94 & 0.23\tablenotemark{b} & 2.3 & 4, 7, 8, 10 \\
Parenago 2494 & Trapezium & 0.71 & 19.48 & 0.4& $>$2.1 & 4, 7, 8, 11\\
Parenago 2486 & Trapezium & 0.96 & 5.19 & 0.9& 1.4 & 4, 7, 8, 11\\
RX J0541.4$-$0324 & Orion & 0.67 & 4.99 & 0.25 & 1.9 & 1 \\
GG Ori\tablenotemark{c} & Orion & 1.00 & 6.63 & ... & 2.342 & 12 \\
W134 & NGC 2264 & 0.96 & 6.35 & ... & 2.0 & 8, 13 \\
RS Cha\tablenotemark{c} & Chameleon & 0.98 & 1.7 & ... & 1.86 & 2 \\
HD 98800 B & TW Hya & 0.83 & 315.15 & ... & 0.87 & 14, 15 \\
NTTS 155913$-$2233 & Sco-Cen& 0.63& 2.42 & 0.39\tablenotemark{b} & 1.0 & 4, 7, 10, 16 \\
RX J1603.8$-$3938 & Lupus & 0.93 & 7.56 & ... & 1.0 & 17 \\
NTTS 160905$-$1859 & Sco-Cen & 0.18 & 10.4 & 0.24\tablenotemark{b} & 1.2 & 4, 7, 10 \\
NTTS 162814$-$2427 & $\rho$ Oph & 0.92 & 35.95 & ... & 1.26 & 2 \\
AK Sco & Sco-Cen ? & 0.99 & 13.61 & $\sim$ 1 & 1.6 & 4, 7, 8 \\
HD 155555 & isolated  & 0.93 & 1.68 & 0.8 & 1.1 & 4, 7, 8, 11\\
V4046 Sgr & isolated  & 0.94 & 2.42 & ... & 0.86 & 18 \\
TY CrA\tablenotemark{c}    & Corona Australis& 0.52 & 2.89 & ... & 3.16 & 2 \\
EK Cep\tablenotemark{c}    & isolated & 0.55 & 4.43 & ... & 2.03 & 2 \\
\enddata
\tablenotetext{a}{References are for data used to estimate
M$_1$, not for discovery papers.}
\tablenotetext{b}{Flux ratio is average of a range of values
determined from use of several $H$ band templates in the 
cross-correlation analysis.  See \S 4.}
\tablenotetext{c}{Eclipsing systems.}
\tablerefs{
(1) Covino et al. 2001;
(2) Palla \& Stahler 2001;
(3) Lee et al. 1994; (4) Luhman 2000; (5) Mathieu et al. 1997;
(6) Steffen et al. 2001;
(7) Mathieu 1994; (8) Palla \& Stahler 1999;
(9) Covino et al. 2000;
(10) this work; (11) White \& Ghez 2001; (12) Torres et al. 2000;
(13) Padgett \& Stapelfeldt 1994;
(14) Torres et al. 1995; (15) Prato et al. 2001;
(16) Walter et al. 1994; (17) Guenther et al. 2001;
(18) Quast et al. 2001.}

\end{deluxetable}

\clearpage

\pagestyle{empty}

\begin{deluxetable}{lcccccc}
\tablewidth{0pt}
\tablecaption{Detailed Sample Properties \label{tbl-5}}
\tablehead{
\colhead{$ $} & \colhead{$ $} & \colhead{M$_1$\tablenotemark{b}} & \colhead{M$_2$\tablenotemark{c}} & \colhead{M$_{2_{HRD}}$} & \colhead{f(M)\tablenotemark{d}} & \colhead{M$_{2_{min}}$}\\
\colhead{Object} & \colhead{q\tablenotemark{a}} & \colhead{(M$_{\odot}$)}& \colhead{(M$_{\odot}$)} & \colhead{(M$_{\odot}$)} & \colhead{(M$_{\odot}$)}&\colhead{(M$_{\odot}$)}}
\startdata
Parenago 1771&0.34$\pm$0.03 &1.4$\pm$0.5 &0.48$\pm$0.18 &0.34$\pm$0.17 & 0.024 &0.43$\pm$0.10\\
Parenago 1925&0.31$\pm$0.04 &2.3$\pm$0.6 &0.71$\pm$0.21 &0.15$\pm$0.22 & 0.011 &0.44$\pm$0.08\\
NTTS 155913$-$2233 &0.63$\pm$0.02 &1.0$\pm$0.2 &0.63$\pm$0.13 &0.34$\pm$0.20 & 0.064 &0.53$\pm$0.06 \\
NTTS 160905$-$1859 &0.18$\pm$0.01 &1.2$\pm$0.2 &0.22$\pm$0.04 &0.16$\pm$0.21 & 0.0042 &0.20$\pm$0.02 \\
\enddata

\tablenotetext{a}{From Table 3.}
\tablenotetext{b}{From Table 4.}
\tablenotetext{c}{~q$\times$M$_1$}
\tablenotetext{d}{From \citet{mat94}.}

\end{deluxetable}

\end{document}